\documentclass[showpacs,preprintnumbers,amsmath,amssymb,superscriptaddress]{revtex4}
\usepackage{epsfig}
\usepackage{psfrag}
\usepackage{amsfonts}
\usepackage{graphicx}
\usepackage{dcolumn}
\usepackage{bm}
\usepackage{xcolor}
\usepackage{amsmath}
\begin{document}


\title{Magnetized neutral 2SC color superconductivity and possible origin of the inner magnetic field of magnetars }

\author{Shuai Yuan}

\author{Bo  Feng}
\affiliation{
School of Physics, Huazhong University of Science and Technology, Wuhan 430074, China
}

\author{Efrain J. Ferrer}
\affiliation{Dept. of Physics and Astronomy, Univ. of Texas Rio Grande Valley, 1201 West University Dr., Edinburg, TX 78539, USA}
\affiliation{Institut d’Estudis Espacials de Catalunya (IEEC)
08860 Castelldefels (Barcelona), Spain} 

\author{Alejandro Pinero}
\affiliation{Dept. of Physics and Astronomy, Univ. of Texas Rio Grande Valley, 1201 West University Dr., Edinburg, TX 78539, USA}


\date{\today}

\begin{abstract}
In this paper the neutral 2SC phase of color superconductivity is investigated in the presence of a magnetic field and for diquark coupling constants and baryonic densities that are expected to characterize neutron stars. Specifically, the behavior of the charged gluons Meissner masses is investigated in the parameter region of interest taking into account, in addition, the contribution of a rotated magnetic field. It is found that up to moderately-high diquark coupling constants the mentioned Meissner masses become tachyonic independently of the applied magnetic-field amplitude, hence signalizing the chromomagnetic instability of this phase. To remove the instability, the restructuring of the system ground state is proposed, which now will be formed by vortices of  the rotated charged gluons. These vortices boost the applied magnetic field having the most significant increase for relatively low applied magnetic fields. Finally, considering that with the stellar rotational frequency observed for magnetars a field of the order of $10^8$ G can be generated by dynamo effect, we show that by the boosting effect just described the field can be amplified to $10^{17}$ G that is in the range of inner core fields expected for magnetars. Thus, we conclude that the described mechanism could be the one responsible for the large fields characterizing  magnetars if the cores of these compact objects are formed by neutral 2SC matter.
\end{abstract}

\maketitle


\section{Introduction}

It is of great interest to the nuclear physics and astrophysics communities to elucidate what are the physical characteristics of the superdense matter that forms the interior of neutron stars (NS).
In this context, to understand the effect of a strong magnetic field on that medium is essential due to the fact that magnetic fields of different intensities populate the entire landscape of these compact objects. The surface magnetic fields of some radio pulsars are of the order  of $10^{8}-10^{12}$ G \cite{NS-B-1}. There are even some special compact objects called magnetars \cite{NS-B-2} whose surface magnetic fields are of the order of $10^{14}-10^{15}$ G. Those values have been inferred from spectroscopic and spin-down studies of soft-gamma ray repeaters (SGRs) and anomalous x-ray pulsars (AXPs). In addition, we should take into account that the inner core magnetic fields of magnetars can be even larger, which follows from the magnetic-field flux conservation in stellar media possessing very large electric conductivities. By using different arguments the inner fields have been estimated to range from $10^{17}$  to $10^{20}$ G \cite{magnetizedfermions, Theor-estimations}. The fact that the stellar inner medium will be under the effect of such strong magnetic fields, has motivated many works focused on the study of the equation of state of magnetized NS (see \cite{insignificance} and references there). On the other hand, to understand what originates the tremendous magnetic fields that characterizes magnetars remains as a question under investigation.

Quark matter under sufficiently cold and dense conditions is expected to be realized in the interior of massive NS \cite{Weber, Quark-stars}. Under such conditions, the existence of a color superconducting (CS) phase is unavoidable. This phase is characterized by the formation of Cooper pairs of quarks due to the attractive interaction originating either from the perturbative one-gluon exchange in the color antisymmetric channel at ultrahigh baryon density or from the instanton effects at moderate density.  When the masses of the $u$, $d$ and $s$ quarks can be ignored compared to the baryon chemical potential and all three flavors take part in the pairing, the ground state is characterized by a spin-zero CS condensate giving rise to what is called as the Color-Flavor-Locked (CFL) phase \cite{CFL}. This phase is the well-established ground state at asymptotic high density and low temperature. On the other hand, at moderate densities, where it is expected that the quark interaction is stronger, if the density is high enough to release quarks from inside the hadrons and sufficiently low to decouple the $s$ quark dynamics, the pairing involves only two light flavors and the ground state is a two-flavor superconducting (2SC) phase \cite{Igor-Rischke}.

In investigating compact stars, there is another physical factor to be considered: the electric neutrality of the medium (given by $\partial\Omega / \partial \mu_e=0$, with $\Omega$ the system thermodynamic potential and $\mu_e$ the electric chemical potential), which can also affect the CS phase specially at moderate densities where it is expected that the so-called 2SC phase can be realized. Some authors have already investigated the neutral 2SC phase in the presence of a magnetic field \cite{Jaikumar-1}-\cite{Mishra}. Those works had been constrained  to study the evolution of the quark gap $\Delta$ and constituent quark mass $m$ with increasing densities and magnetic fields. Nevertheless, the effect of an applied magnetic field on the gluon sector  of such an electrically neutral system has been ignored up to now. 

Considering that quark matter in the interior of NS must be neutral with respect to both electric and color charges and in $\beta$ equilibrium, (i.e., the weak beta-equilibrium condition $\mu_u+\mu_e-\mu_{\nu_e}=\mu_d$ should be satisfied, with $\mu_u$ the chemical potential of the $u$ quark, $\mu_d$ the one for the $d$ quark, $\mu_e$ the electron chemical potential and $\mu_{\nu_e}$ that of the electron neutrinos, which will be here ignored considering stars in the post neutrino-emission epoch.)
it was found \cite{Igor} that the pairing dynamics within the 2SC phase results affected due to the mismatch between the Fermi spheres of quarks with different flavors. As a  consequence, gapless modes in the spectrum of quasiparticles appear when the mismatch $\delta \mu_e = \mu_e/2$ satisfies the condition $0<\Delta/\delta\mu_e<1$. This unusual phase is called the gapless 2SC phase (g2SC).  
Further calculations \cite{Huang} revealed the other important consequence that the square of the Meissner mass of some gluons are negative signaling an instability, which was named the chromomagnetic instability.  It is worth to emphasize that the gapless superconductivity itself is not the reason for such an instability since, for example,  in the gapped 2SC phase the squared Meissner mass of some gluons are also negative when $1<\Delta/\delta\mu_e<\sqrt{2}$. The chromomagnetic instability poses a major challenge in the studies of CS phases with quark pairing mismatch.  Although a consensual conclusion to remove such an instability hasn't been reached, possible alternative solutions already exist in the literature. 
In this context, inhomogeneous solutions seem to play an important role. Some of them spontaneously break translational invariance \cite{Reddy}-\cite{Hashimoto} through inhomogeneous quark-quark condensates which involves a momentum-dependent gap. These kind of inhomogeneous CS phases are based on the idea of Larkin and Ovchinnikov \cite{Larkin} and Fulde and Ferrell \cite{Ferrell} originally applied to condensed matter. Another inhomogeneous phase that breaks rotational invariance has been also considered where gluon vortices are induced together with the spontaneous generation of a magnetic field \cite{Ferrer-Vortices}. This, second scenario is motivated by the solution to the so-called "zero-mode problem" that takes place for a charged spin-one field in the presence of a magnetic field larger than the field square Meissner mass \cite{Yang-Mills}-\cite{String Theory}.

In the present work, we will show that in the presence of a magnetic field the unstable neutral 2SC phase will be stabilized by restructuring its ground state by the formation of a gluon vortex condensate as follows from the well-known solution of the ”zero-mode problem,” which has been already found in different contexts as for Yang-Mills fields in \cite{Yang-Mills}, for
the $W^\pm_\mu$ bosons of the electroweak theory in \cite{ W-EW}, for the charged gluons of the MCFL phase of CS in \cite{MCFL-Vortices} and
even for higher-spin fields in the context of string theory in \cite{String Theory}. In the magnetized neutral 2SC system the gluon vortex solution will be even reinforced for certain density regions in the presence of a magnetic field as we will show in this work.

Something else to notice is that an important characteristic of spin-zero CS phases is the lack of Meissner effect in contrast to the conventional superconductivity. Although the original electromagnetic $U(1)_{em}$ symmetry is broken by the formation of quark Cooper pairs, a residual $\widetilde{U}(1)$ symmetry still remains. The massless gauge field associated with this symmetry in the 2SC phase is given by the linear combination of the conventional photon field  $A_\mu$ and the 8th-gluon field $G^8_\mu$  \cite{Rotated Fields} as $\widetilde{A}_{\mu}= \cos \theta A_{\mu}-\sin \theta G^8_{\mu}$. The field $\widetilde{A}_\mu$ plays the role of a long-range in-medium electromagnetic field, which is also called the rotated electromagnetic field, while the orthogonal combination $\widetilde{G}_{\mu}^8= \sin\theta A_{\mu}+\cos\theta G^8_{\mu}$ is massive. The mixing angle depends on the coupling constants through the relation $\tan \theta=e/\sqrt{3}g$, with $g$ and $e$ the strong and electromagnetic coupling constants, respectively. In this paper we will investigate the effect of the rotated magnetic field associated with  $\widetilde{A}_\mu$ on neutral 2SC matter and especially on the Meissner masses of gluons that acquire rotated electric charges, $G^{(4,5,6,7)}_\mu$, and which are the ones associated with the tachyonic modes in this medium. The mechanism that serves to remove in this context the chromomagnetic instability is due to the formation of gluonic vortices together with the generation of a magnetic field. Thus, we will also show how much the applied magnetic field can be boosted by the generation of the gluon vortices that can be induced at the moderately high baryon density expected to be reached in NS.
Finally, based on those results, one of the main outcomes of this paper will be to present a natural mechanism to explain the possible origin of the strong magnetic fields exhibited by magnetars.

The rest of the paper is organized as follows: In Section 2, we review the NJL formulation of the 2SC neutral phase of CS in the presence of a uniform magnetic field. The results for the gap, $\Delta$, and electric chemical potential, $\mu_e$, found in this section will be used then in the subsequent sections. In Section 3, the Meissner mass of the rotationally charged gluons is calculated for different values of the diquark coupling constant in the range of baryonic densities that are considered to be of interest for NS. It will be shown that for coupling constants not extremely high, the Meissner mass exhibits a tachyonic behavior at moderately high magnetic fields in the whole range of densities under consideration. This is what has been called the chromomagnetic instability phenomenon. It will be discussed how this instability is removed by redefining the system ground state through the formation of gluon vortices. A peculiar effect generated by these vortices is the amplification of the magnitude of  the applied magnetic field. By numerical calculations we will see that small fields up to $10^{15}$ G will be boosted in 2 orders of magnitude. This effect will be used then in Section 4 as a mechanism to produce the high strength that the magnetar magnetic fields can reach in its interior. In Section 5 we present the concluding remarks.

\section{Cooper pair condensate at ${\bm B}\ne 0$ in the neutral medium}

We consider in this paper a two-flavor NJL model of massless quarks in the presence of an external constant and uniform rotated magnetic field (from now on we are going to ignore the term "rotated" but it will be indicated by a wavy-hat notation on top of the magnitudes). The Lagrangian reads
\begin{align}
{\cal L}=\bar\psi\left[i\gamma^\mu(\partial_\mu+i\tilde{e}\tilde{Q}\tilde{A}_\mu)+\mu\gamma^0\right]\psi+G_S\left[(\bar\psi \psi)^2+(\bar\psi i\gamma^5{\tau^3}\psi)^2\right]+G_D\left(\bar\psi i\gamma^5T^2\tau^2\psi^C \right)\left(\bar\psi^Ci\gamma^5T^2\tau^2\psi\right),\label{Lagrangian}
\end{align}
where  $\psi^T=(u,d)$ is the quark doublet and $\psi^C=C\bar\psi^T$  is its charge conjugate with $C=i\gamma^2\gamma^0$  the charge conjugation operator, and $G_S$ and $G_D$ the coupling constants of the scalar and diquark channels of the four-fermion interaction theory respectively. Here, $T^2$ is the second Gell-Mann matrix in color space and $\tau^{2,3}$ are Pauli  matrices in flavor space.   $\tilde{Q}$ is the rotated charge matrix defined as 
\begin{equation}
\tilde{Q}=Q_f\times 1_c-1_f\times \frac{T_c^8}{2\sqrt{3}}.
\end{equation}
in units of $\tilde{e}=\frac{\sqrt{3}ge}{\sqrt{3g^2+e^2}}>0$.  Here $Q_f={\rm diag}(2/3, -1/3)$ is the original electromagnetic charge matrix for quarks and $T_c^8={\rm diag}(1/\sqrt{3}, 1/\sqrt{3},-2/\sqrt{3})$ is the 8th Gell-Mann matrix. To be explicit, the $\tilde{Q}$ charges for different quarks are given in Table I and for different gluons in Table II.
 \begin{table}[htbp]
	\centering
	\caption{$\tilde{Q}$ charge for different quarks}
	\begin{tabular}{cccccc}
		\hline
		$u_r$\ \ \ \ & $u_g$\ \ \ \ & $u_b$\ \ \ \ & $d_r$\ \ \ \ & $d_g$\ \ \ \ & $d_b$  \\
		\hline
		$\frac{1}{2}$\ \ \ \ & $\frac{1}{2}$\ \ \ \ & 1\ \ \ \ & $-\frac{1}{2}$\ \ \ \ & $-\frac{1}{2}$\ \ \ \ & 0  \\
		\hline
	\end{tabular}
\end{table}
In this convention, the only diquark pairs are $u_rd_g$, and $u_gd_r$. 
These pairs are neutral with respect to $\tilde{Q}$. This is the reason why there is no Meissner effect  for $\tilde{B}$ due to the absence of a charged ground state.

For the external magnetic field we impose the Landau gauge, in which the electromagnetic potential is given by $\tilde{A}^\mu=(0,0,\tilde{B}x,0)$. This corresponds to a constant and uniform magnetic field $\tilde{B}$  in the positive $z$ direction. Note that, in the 2SC phase the linear combinations of the gluon fields 
\begin{equation}
G_\mu^\pm=\frac{1}{\sqrt{2}}(G_\mu^4\mp iG_\mu^5), \ \ \ \ \ \ H_\mu^\pm=\frac{1}{\sqrt{2}}(G_\mu^6\mp iG_\mu^7)
\end{equation}
carry $\tilde{Q}$ charges, while the remaining gluon fields are neutral with respect to $\tilde{Q}$.
\begin{table}[htbp]
	\centering
	\caption{$\tilde{Q}$ charge for different gluons}
	\begin{tabular}{cccccccc}
		\hline
		$G_\mu^1$\ \ \ \ & $G_\mu^2$\ \ \ \ & $G_\mu^3$\ \ \ \ & $G_\mu^+$\ \ \ \ & $G_\mu^-$\ \ \ \ & $H_\mu^+$\ \ \ \ & $H_\mu^-$ \ \ \ \ & ${\tilde G}_\mu^8$\\
		\hline
		$0$\ \ \ \ & $0$\ \ \ \ & $0$\ \ \ \ & $\frac{1}{2}$\ \ \ \ & $-\frac{1}{2}$\ \ \ \ & $\frac{1}{2}$\ \ \ \ & $-\frac{1}{2}$\ \ \ \ &0\\
		\hline
	\end{tabular}
\end{table}

The quark chemical potentials related with electrical and color neutralities, as well as $\beta$ equilibrium can be expressed by the following matrix in color-flavor space: 
\begin{equation}
\mu_{ij,\alpha\beta}=\left[\mu\delta_{ij}-\mu_e(Q_f)_{ij}\right]\delta_{\alpha\beta}+\frac{2}{\sqrt{3}}\mu_8\delta_{ij}(T^8)_{\alpha\beta},
\end{equation}
with components that can be more explicitly written for each quark as
\begin{equation}\label{u-r}
\mu_{u_r}=\mu_{u_g}=\bar\mu-\delta\mu_e
\end{equation}
\begin{equation}\label{d-r}
\mu_{d_r}=\mu_{d_g}=\bar\mu+\delta\mu_e
\end{equation}
\begin{equation}\label{u-b}
\mu_{u_b}=\bar\mu-\delta\mu_e-\delta\mu_8
\end{equation}
\begin{equation}\label{d-b}
\mu_{d_b}=\bar\mu+\delta\mu_e-\delta\mu_8
\end{equation}
The notations  $\bar{\mu}=\mu-\frac{1}{6}\mu_{e}+\frac{1}{3}\mu_{8}$,  $\delta\mu_e=\frac{1}{2}\mu_{e}$ and $\delta\mu_8=\mu_{8}$ have been used. 
Notice that with the definitions (\ref{u-r})-(\ref{d-b}) the weak beta-equilibrium condition $\mu_u+\mu_e-\mu_{\nu_e}=\mu_d$ is satisfied for each quark color.
The chemical potentials $\mu_e$ and $\mu_8$ are dynamical parameters that have to be determined from the electrical and color neutrality equations. The color chemical potential $\mu_8$ is usually very small compared with the other chemical potentials $\mu_e$  and $\mu$ \cite{mu-8}. We, therefore, will assume $\mu_8=0$ in the following calculations since as has been found by other authors it will not significantly change the results.
 
The partition function reads
\begin{align}
{\cal Z}=\int{\cal D}\bar\psi{\cal D}\psi\exp({{\cal S}_{\rm eff}})
\end{align}
with ${\cal S}_{\rm eff}$ the effective action in Euclidean space given by

\begin{align} 
{\cal S}_{\rm eff} =& \int_0^\beta d\tau\int d^3{\bf x}\left\{ \bar\psi\left[-\gamma^0\frac{\partial}{\partial \tau}+\gamma\cdot(i\nabla+\tilde{e}\tilde{Q}\tilde{\bf A})+\mu\gamma^0\right]\psi\right. \nonumber 
\\ 
&+\left.\frac{1}{2}\Delta\bar\psi i\gamma^5T^2\tau^2\bar\psi^C+\frac{1}{2}\Delta^*{\bar\psi}^Ci\gamma^5T^2\tau^2\psi-\frac{|\Delta|^2}{4G_D}\right\}
\end{align}
which is obtained from Eq. (\ref{Lagrangian}) by assuming $\Delta=2G_D\langle {\bar\psi}^Ci\gamma^5T^2\tau^2\psi \rangle$ and making the mean field approximation.  Note that we are not considering the chiral condensation. From previous results  \cite{Jaikumar-2}, we see that for the density region $\mu > 320$ MeV the chiral condensate already vanished. Thus, in our calculations we will consider only the region with $\mu > 320$ MeV, which on the other hand is the region where $\Delta \neq 0$ \cite{Jaikumar-2}.

Introducing the Nambu-Gorkov spinors,
\begin{align}
&\Psi_{({\tilde Q})}=\left(\begin{array}{c}\psi_{({\tilde Q})}\\ \psi^{C}_{(-{\tilde Q})}\end{array}\right),\quad\bar{\Psi}_{({\tilde Q})}=\left(\begin{array}{cc}\bar{\psi}_{({\tilde Q})}&\bar{\psi}^{C}_{(-{\tilde Q})}\end{array}\right)
\end{align}
where $\psi_{({\tilde Q})}=\Omega_{({\tilde Q})}\psi$ with charge projectors for the field representation $\psi^T=(u_r,u_g,u_b,d_r,d_g,d_b)$ given by
\begin{equation}
\Omega_{(0)}={\rm diag}(0,0,0,0,0,1)
\end{equation}
\begin{equation}
\Omega_{(\frac{1}{2})}={\rm diag}(1,1,0,0,0,0)
\end{equation}
\begin{equation}
\Omega_{(-\frac{1}{2})}={\rm diag}(0,0,0,1,1,0)
\end{equation}
\begin{equation}
\Omega_{(1)}={\rm diag}(0,0,1,0,0,0)
\end{equation}
in color and flavor space. Those projectors satisfy
\begin{equation}
\Omega_{\eta}\Omega_{\eta^\prime}=\delta_{\eta\eta^\prime}\Omega_{\eta}, \ \ \ \ \ \ \ \ \eta, \eta^\prime=0,\pm\frac{1}{2}, 1
\end{equation}
and
\begin{equation}
\sum_\eta\Omega_{\eta}=1
\end{equation}
The effective action can be written as
\begin{equation}
{\cal S}_{\rm eff}=\int_0^\beta d\tau\int d^3{\bf x}\left[\frac{1}{2}\sum_{\tilde Q}{\bar\Psi}_{({\tilde Q})}S_{({\tilde Q})}^{-1}\Psi_{({\tilde Q})}-\frac{\Delta^{2}}{4G}\right]         
\end{equation}
with
\begin{equation}
S_{({\tilde Q})}^{-1}=\left(\begin{array}{cc}
[G^+_{({\tilde Q})0}]^{-1} & \Phi^-_{({\tilde Q})}\\
\Phi^+_{({\tilde Q})} & [G^-_{({\tilde Q})0}]^{-1}\\
\end{array}\right)
\end{equation}
The diagonal elements are given by
\begin{equation}
[G^\pm_{({\frac{1}{2}})0}]^{-1}=-\gamma^0\frac{\partial}{\partial\tau}+\gamma\cdot\left( i\nabla+\frac{1}{2}|{\tilde e}{\tilde {\bf A}}|\right)\pm({\bar\mu}\mp\delta\mu_e)\gamma^{0}
\end{equation}
and
\begin{equation}
[G^\pm_{(-{\frac{1}{2}})0}]^{-1}=-\gamma^0\frac{\partial}{\partial\tau}+\gamma\cdot\left( i\nabla-\frac{1}{2}|{\tilde e}{\tilde {\bf A}}|\right)\pm({\bar\mu}\pm\delta\mu_e)\gamma^{0}
\end{equation}
\begin{equation}
[G^\pm_{(0)0}]^{-1}=-\gamma^0\frac{\partial}{\partial\tau}+\gamma\cdot i\nabla\pm({\bar\mu}+\delta\mu_e-\delta\mu_8)\gamma^{0}
\end{equation}
\begin{equation}
[G^\pm_{(1)0}]^{-1}=-\gamma^0\frac{\partial}{\partial\tau}+\gamma\cdot\left( i\nabla+|{\tilde e}{\tilde {\bf A}}|\right)\pm({\bar\mu}-\delta\mu_e-\delta\mu_8)\gamma^{0}
\end{equation}
The off-diagonal elements are given by
\begin{equation}
\Phi^+_{({\tilde Q})}=0, \ \ \ \ \ {\tilde Q}=0,1
\end{equation}
\begin{equation}
\Phi^+_{(\frac{1}{2})}=-\Phi^+_{(-\frac{1}{2})}=\left(\begin{array}{cc}
0 & \Delta i\gamma^5\\
-\Delta i\gamma^5 & 0\\
\end{array}\right)
\end{equation}
and
\begin{equation}
\Phi^-_{({\tilde Q})}=\gamma^0\left[\Phi^+_{({\tilde Q})}\right]^\dagger\gamma^0
\end{equation}

\begin{figure}
	\includegraphics[height=12cm]{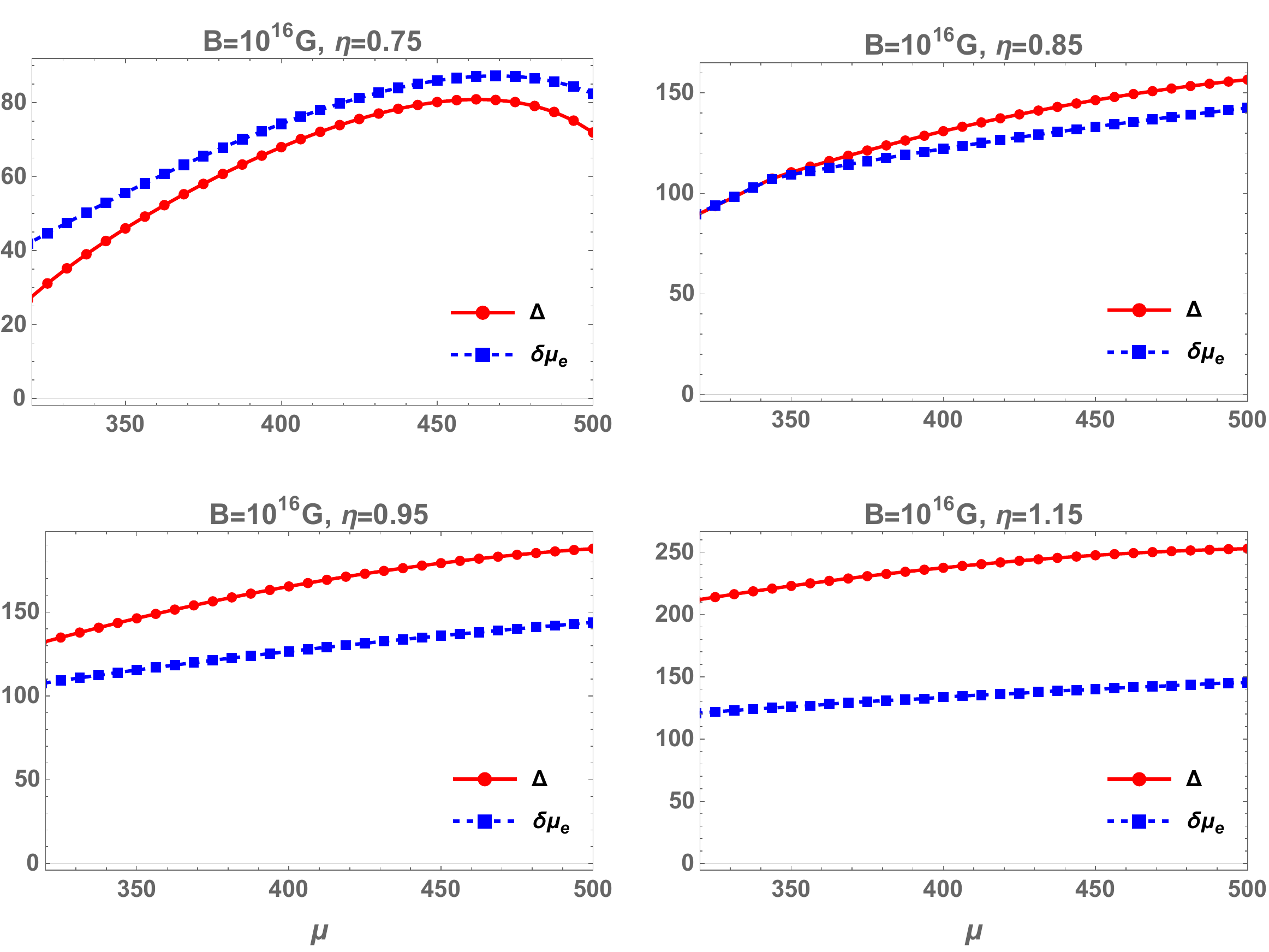}
	\caption{The solutions of the gap $\Delta$ and mismatch $\delta \mu_e$ versus baryonic chemical potential for different diquark couplings at a magnetic-field value of $10^{16}$G.} \label{Fig_1}
\end{figure}

\begin{figure}
	\includegraphics[height=12cm]{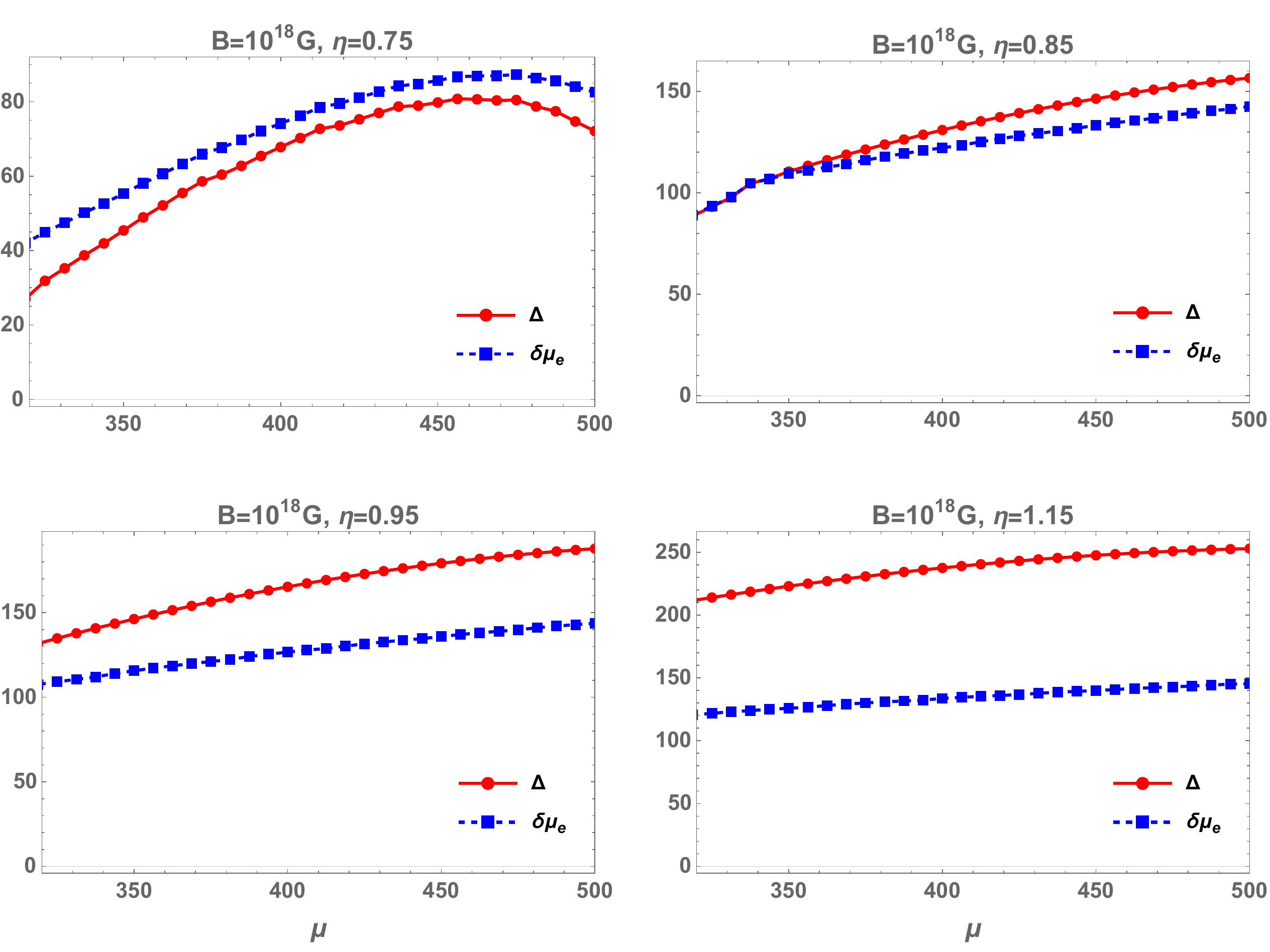}
	\caption{The solutions of the gap $\Delta$ and mismatch $\delta \mu_e$ versus baryonic chemical potential for different diquark couplings at a magnetic-field value of $10^{18}$ G.}\label{Fig_2}
\end{figure}

Thus, we obtain the thermodynamical potential
\begin{align}
\nonumber\Omega &=-\frac{1}{\beta V}\ln{\cal Z}-\frac{1}{12\pi^2}\left(\mu_e^4+2\pi^2T^2\mu^2+\frac{7\pi^4}{15}T^4\right)\\
&=\frac{\Delta^{2}}{4G}-\sum_{({\tilde Q})}\ln\det S_{({\tilde Q})}^{-1}-\frac{1}{12\pi^2}\left(\mu_e^4+2\pi^2T^2\mu^2+\frac{7\pi^4}{15}T^4\right)
\end{align}
with the contribution in the parentheses coming from the electrons, which must be introduced to maintain the system neutrality and $\beta$ equilibrium. 

Carrying out the sum in Matsubara frequencies,  we finally have
\begin{align}
\nonumber \Omega=&-\frac{1}{12\pi^2}\left(\mu_e^4+2\pi^2T^2\mu^2+\frac{7\pi^4}{15}T^4\right)+\frac{\Delta^2}{4G}-\int\frac{d^3{\bf p}}{(2\pi)^3}\sum_{\sigma=\pm}\left(E_{(0)}^\sigma+2T\ln\left[1+e^{-\frac{ E_{(0)}^\sigma}{T}}\right]\right)\\
&-\frac{2|{\tilde e}{\tilde Q}{\tilde {B}}|}{(2\pi)^2}\int dp_3\sum_{l=0}^\infty(1-\frac{1}{2}\delta_{l,0})\sum_{{\tilde Q}=1, \pm\frac{1}{2}}\sum_{\sigma, \xi=\pm}\left(E_{({\tilde Q})\xi}^\sigma+2T\ln\left[1+e^{-\frac{E_{(\tilde Q)\xi}^\sigma}{T}}\right]\right),
\end{align}
where $E_{\tilde Q}$ are the spectrum of the quasiparticles given with their corresponding degeneracies by
\begin{align}
E_{(0)}^\pm=&\pm\mu_{d_b}+|{\bf p}|\ \ \ \ \ \ \ \   \   (\times 2)\\
E_{(1)}^\pm=&\pm\mu_{u_b}+|{\bar{\bf p}}^{(+)}|\ \ \ \ \  (\times 2)\\
E_{(\frac{1}{2})\pm}^\pm=&\pm\delta\mu_e+\sqrt{(|{\bar{\bf p}}_{(\frac{1}{2})}|\pm\bar{\mu})^{2}+\Delta^{2}}\\
E_{(-\frac{1}{2})\pm}^\pm=&\pm\delta\mu_e+\sqrt{(|{\bar{\bf p}}_{(-\frac{1}{2})}|\pm\bar{\mu})^{2}+\Delta^{2}}
\end{align}  
 with ${\bar{\bf p}}_{{(\tilde Q)}}=(0, {\rm sign}({\tilde Q})\sqrt{2|{\tilde e}{\tilde Q}{\tilde {B}}|}, p^3)$ and $E^\sigma_{(1)\xi}=E^\sigma_{(1)}$ understood. In the present paper, we shall treat $\mu_e$ as a dynamical parameter whose value should be determined by minimizing the thermodynamical potential together with the gap equation for $\Delta$
 \begin{equation}\label{gap-eq}
 \frac{\partial\Omega}{\partial\Delta}=\frac{\partial\Omega}{\partial\mu_e}=0,
 \end{equation}
which will be solved numerically. The NJL model parameters, i.e., the three-momentum cutoff $\Lambda$ and the coupling constant $G_S$, are chosen to be $\Lambda=653.3 \ {\rm MeV}$ and $G_S\Lambda^2=2.14$, which was fixed by fitting the pion mass and its decay constant measured in experiment \cite{Buballa-Rep}. However, the coupling constant $G_D$ in the diquark channel was not fixed by any phenomenologies. We will simply use the convention that $G_D$ was proportional to $G_S$, i.e., $G_D=\eta G_S$, with values $\eta=0.75, 0.85, 0.95$ and $1.15$.

In Figs. \ref{Fig_1} and \ref{Fig_2}, we present for two different values of magnetic fields the solutions of the minimum equations (\ref{gap-eq}) in the range of baryonic chemical potentials from 320 to 500  MeV, which are expected to correspond to the typical density region of compact stars. 
Comparing Figs. 1 and 2 we notice that the magnetic field does not significantly affect the behavior of $\Delta$ and $\delta \mu_e$ in the parameter range under consideration. As expected, the coupling strength $\eta$ affects significantly the value of the gap $\Delta$ producing a significant increase. The electric chemical potential, on the other hand, also increases with $\eta$. This is indicating that as more quarks having a positive conventional electric charge condense in Cooper pairs, the system  needs a larger number of electrons to keep its electric neutrality.

\begin{figure}
	\includegraphics[height=12cm]{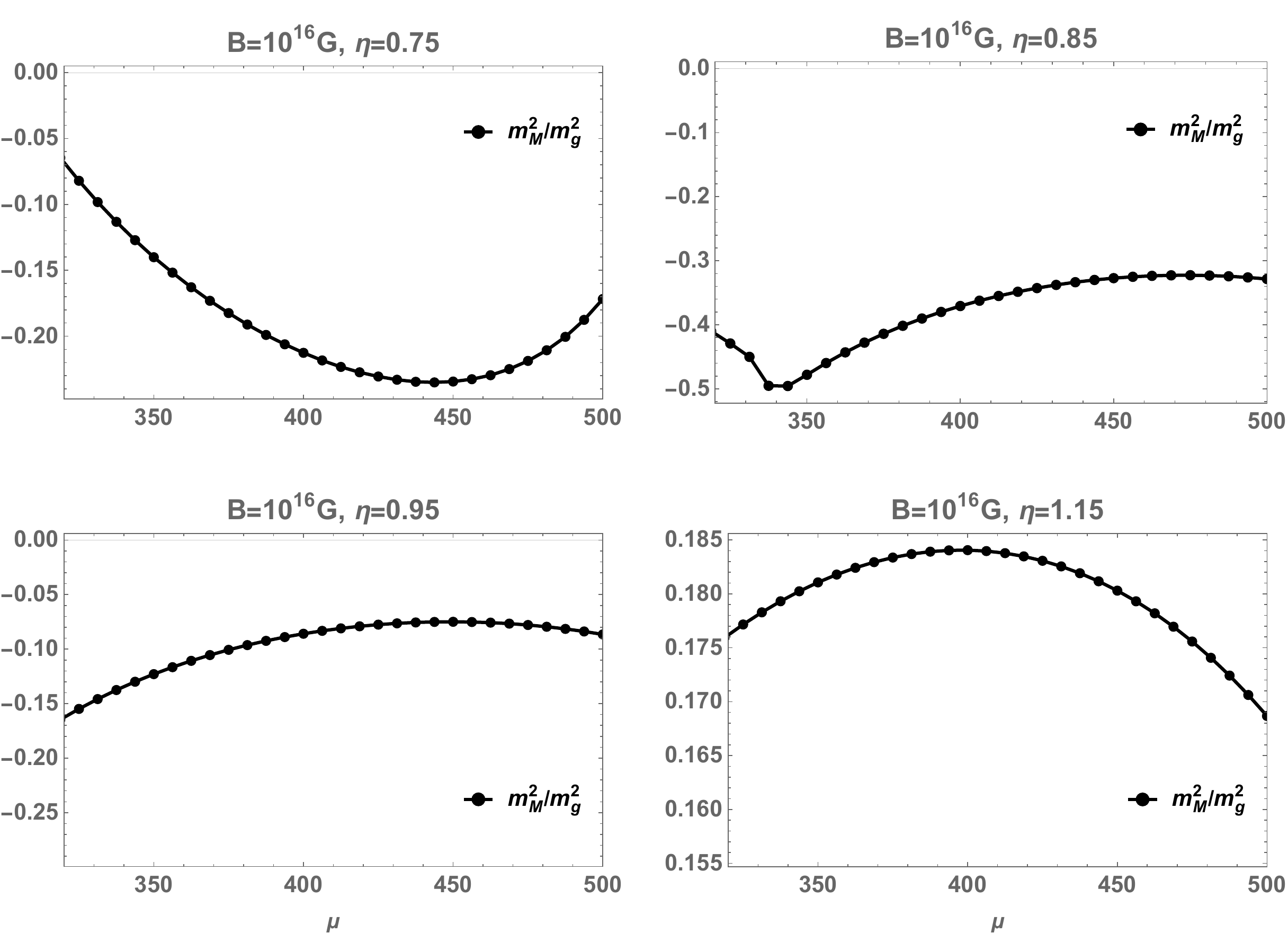}
	\caption{The ratio of the squared Meissner mass of charged gluons to the magnetic mass for different diquark couplings at a magnetic-field value of $10^{16}$ G. } \label{Cromo-M-Inst-1}
\end{figure}

\begin{figure}
	\includegraphics[height=12cm]{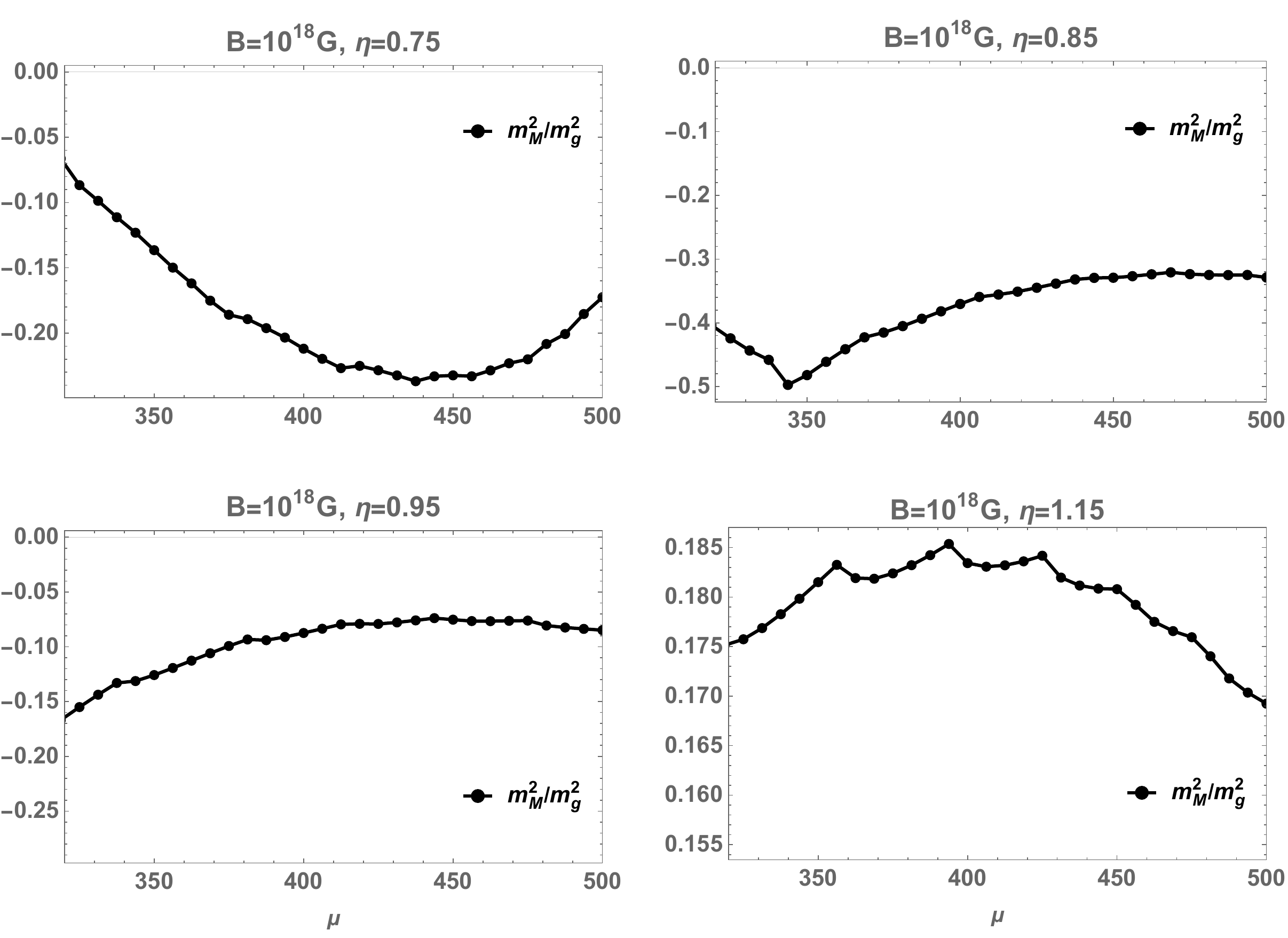}
	\caption{The ratio of the squared Meissner mass of charged gluons to the magnetic mass for different diquark couplings at a magnetic-field value of $10^{18}$ G.} \label{Cromo-M-Inst-2}
\end{figure}

\section{The removal of the Chromomagnetic instability of neutral 2SC matter at ${\bm B}\ne 0$ as a mechanism to boost a magnetic field}

The gluon screening properties in the neutral 2SC phase can be studied by calculating the gluon polarization tensors considered in detail in Ref. \cite{Huang}. In principle, for sufficiently high magnetic fields the polarization tensor of charged particles can change their Lorentz tensorial structure showing a space anisotropy determined by the field direction \cite{Pi_Anisotropy}. Nevertheless, for astrophysical applications to compact objects since $eB \ll \mu^2$, the anisotropy in the Lorentz structure of the polarization operator becomes a second-order effect that can be neglected, and, hence we can keep the polarization operator structure characteristic of a dense medium at zero field reducing the effect of the magnetic field to its dependence on $\Delta$ and $\mu_e$.

 In  \cite{Huang} it was found that while the Meissner masses of the three gluons that correspond to the unbroken $SU(2)_c$ subgroup are vanishing as expected, the charged gluons that correspond to the broken generators of the $SU(3)_c$ group could be imaginary in some parameter regions. The explicit Meissner masses for the  charged  and the rotated $8$th gluon fields are given,  respectively, by
 \begin{equation} 
m^2_{M,{\rm charged}}=\frac{2\alpha_s{\bar\mu}^2}{3\pi}\left[\frac{\Delta^2-2(\delta\mu_e)^2}{\Delta^2}+2\frac{\delta\mu_e\sqrt{(\delta\mu_e)^2-\Delta^2}}{\Delta^2}\theta(\delta\mu_e-\Delta)\right]\label{4to7Meissner}
\end{equation}\
 and
 \begin{equation}
m^2_{M,{\tilde 8}{\tilde 8}}=\frac{4(3\alpha_s+\alpha){\bar\mu}^2}{27\pi}\left[1-\frac{\delta\mu_e\theta(\delta\mu_e-\Delta)}{\sqrt{(\delta\mu_e)^2-\Delta^2}}\right]\label{8thMeissner}
\end{equation}
where $\alpha=g^2/4\pi$ and $\alpha_s=e^2/4\pi$.  

We must  also indicate that if the magnetic field is taken into account in the calculation of the polarization operators from where the gluon Meissner masses are obtained, we should expect corrections into Eqs. (\ref{4to7Meissner})-(\ref{8thMeissner}) proportional to ${\tilde e}{\tilde B}/{\bar\mu}^2<<1.$  This is a weak radiative effect that will be of the order of $\alpha_s \tilde{e} \tilde{B}$ and that can be neglected as compared with the tree contribution to the Meissner masses of the order of $\tilde{e}\tilde{B}$ produced by the magnetic-field direct interaction with the charged gluons through the anomalous magnetic moment term  $i{\tilde e}{\tilde f}_{\mu\nu} G_\mu^+G_\nu^-$, as we will see below. Thus, in our calculations the magnetic field will enter in the Meissner masses (\ref{4to7Meissner}) and (\ref{8thMeissner}) only through the dependence of $\Delta$ and $\mu_e$ on this.

From  (\ref{4to7Meissner}) and  (\ref{8thMeissner}),  one can see that the squared Meissner masses of the charged gluons can be negative in the region $0<\Delta/\delta\mu_e<\sqrt{2}$ and for the 8th rotated gluon  in the region $0<\Delta/\delta\mu_e<1$.  The appearance of those negative square Meissner masses is pointing out to the existence of an instability that is called the chromomagnetic instability.

\subsection{Charged gluon Meissner masses versus baryonic chemical potential at ${\bm B}\ne 0$}

In Figs. \ref{Cromo-M-Inst-1} and \ref{Cromo-M-Inst-2}, we plot the ratio of the Meissner mass of charged gluons,  $m_M^2/m_g^2$, with respect to the so-called gluon mass, $m_g^2=4\alpha_s{\bar\mu}^2/3\pi$, versus the baryonic chemical potentials of interest for NS. 
Using different parametrizations for the coupling constant amplitude coefficient $\eta$, the results are shown in Fig.  \ref{Cromo-M-Inst-1} at a moderate magnetic-field value of $10^{16}$ G  and in Fig. \ref{Cromo-M-Inst-2} for a higher value of $10^{18}$ G. 
In both sets of graphs only when the coupling turns very high (i.e. for $\eta=1.15$) does the Meissner masses stop being tachyonic. We call attention to the oscillations in the graphs which are originated by the discontinue change in the occupation of different Landau levels as $\mu$ varies at a constant field. These are the known de Haas-van Alphen oscillations.

Comparing Figs. \ref{Cromo-M-Inst-1} and \ref{Cromo-M-Inst-2}, we also observe that going from moderate to strong magnetic-field values, $m_M^2$ does not significantly change. We also want to call attention to the fact that the Meissner mass does not change monotonically with the baryonic chemical potential since its dependence is through two quantities, $\Delta$ and $\delta \mu_e$, which vary indistinctly with $\mu$.

\subsection{Induced magnetic field in the neutral 2SC phase.}

We should notice that the rotated magnetic field also plays a very peculiar additional role on the Meissner masses of the charged quarks in the color superconducting phase. Since the rotated magnetic field contains a component associated with the 8-th gluon, it contributes to the Meissner masses of the charged gluons through the non-Abelian interaction with the charged gluons. 

This can be seen taking into account the effective action for the charged gluon fields, $G_\mu^\pm$, in the background of the in-medium rotated magnetic field, $\tilde{H}$, taken along the z direction, which is given by \cite{Ferrer-Vortices, MCFL-Vortices}
\begin{align}
\nonumber \Gamma_{\rm eff}^c=&\int d^4x \left\{-\frac{1}{4}({\tilde f}_{\mu\nu})^2-\frac{1}{2}\left({\tilde\Pi}_\mu G_\nu^--{\tilde\Pi}_\nu G_\mu^-\right)^2-\left[\left(m_D^2\delta_{\mu 0}\delta_{\nu 0}+m_M^2\delta_{\mu i}\delta_{\nu i}\right)+i{\tilde e}{\tilde f}_{\mu\nu} \right]G_\mu^+G_\nu^-\right.\\
&+\left.\frac{g^2}{2}\left[(G_\mu^+)^2(G_\nu^-)^2-(G_\mu^+G_\nu^-)^2 \right]+\frac{1}{\lambda}G_\mu^+{\tilde\Pi}_\mu{\tilde\Pi}_\nu G_\nu^-\right\}\label{effectiveaction}
\end{align}
with ${\tilde\Pi}_\mu=\partial_\mu-i{\tilde e}{\tilde A}_\mu$ the covariant derivative, $\tilde{f}_{\mu \nu}=\partial_\mu \tilde{A}_\nu-\partial_\nu \tilde{A}_\mu$, the Debye mass, $m_D$, the Meissner mass $m_M$ and $\lambda$ an arbitrary gauge fixing parameter 
in the 't Hooft gauge. Essentially, the effective action (\ref{effectiveaction}) is the one of a spin-$1$ charged boson field in a magnetic field. Because of the anomalous magnetic moment term  $i{\tilde e}{\tilde f}_{\mu\nu} G_\mu^+G_\nu^-$, even in the case when $m_M^2>0$, one of the charged gluon 
modes becomes imaginary once the field surpasses a critical value, i.e., ${\tilde e}{\tilde B}>{\tilde e}{\tilde B}_c=m_M^2$. This is the well-known "zero-mode problem" for the Yang-Mills fields in the presence of a magnetic field \cite{Ferrer-Vortices}-\cite{String Theory}. The solution to this zero-mode problem is the formation of an inhomogeneous gluon field condensate $\langle G_\mu^\pm\rangle$ together with the amplification of the applied magnetic field.

Replacing in (\ref{effectiveaction}) with the condensate ansatz employed in Ref.  \cite{MCFL-Vortices} where $\langle G_1^-\rangle=-i\langle G_2^-\rangle= G(x,y)$, $\langle G_3^-\rangle=\langle G_0^-\rangle=0$ and $\langle G_i^+\rangle=\langle G_i^-\rangle^*$ we find the system free energy $\cal{F}$.  Then, in the presence of an applied rotated magnetic field $\tilde{H}$ and an induced rotated magnetic field $\tilde{B}$ the Gibbs free energy is obtained as ${\cal G}={\cal F}-\tilde{B}\tilde{H}$, which in our case reads
\begin{equation}\label{E-A-1}
{\cal G}={\cal F}_{0}-2 G^\dagger{\tilde\Pi}^2 G-2(2{\tilde e}{\tilde B}-m_M^2)|G|^2+2g^2|G|^4+\frac{1}{2}{\tilde B}^2-{\tilde H}{\tilde B}
\end{equation}
where ${\cal F}_0$ is the free energy density of the normal phase ($G=0$) at zero applied field. As can be seen from the third term of the rhs of (\ref{E-A-1}) ${\tilde e}\tilde{B}$ enters as a negative square mass. In the case that $m_M^2 > 0$ if ${\tilde e}\tilde{B} >m_M^2/2$ the corresponding gluon mode becomes tachyonic and the ground state of the system has to be modified. In the case of neutral 2SC matter the situation in most of the region of interest at moderate values of baryonic density is even more  pressing since $m_M^2 < 0$. Thus, in this case a magnetic field, no matter how weak it can be, will activate the mechanism (i.e. no critical field is needed).

To find the new ground state where the instability is removed we should minimize the Gibbs free energy with respect to $G$ and $\tilde{B}$. Hence, we find the system of equations
\begin{equation}
-{\tilde\Pi}^2G-(2{\tilde e}{\tilde B}-m_M^2)G+2g^2|G|^2G=0,\label{gluoncondensateequation}
\end{equation}

\begin{equation}
2{\tilde e}|G|^2-{\tilde B}+{\tilde H}=0\label{inducedmagneticfield}
\end{equation}
from where we can find the vortex condensate solution $G$ and the induced magnetic field $\tilde B$. Note that because of the different sign in the applied and induced fields, the field $\tilde B$ can be dynamically induced by the contribution of the gluon anomalous magnetic moment $2{\tilde e}|G|^2$, which comes from the original term in (\ref{effectiveaction}) $i{\tilde e}{\tilde f}_{\mu\nu} G_\mu^+G_\nu^-$. Also see that from (\ref{inducedmagneticfield}) the induced field $\tilde{B}$ is larger than the applied field $\tilde H$. This is the magnetic-field boosting mechanism that can be activated in this circumstance. Our goal from now on is to see how this mechanism can work to produce the strong fields exhibited by magnetars.

\begin{figure}\label{Induced_Field_1}
	\includegraphics[height=8cm]{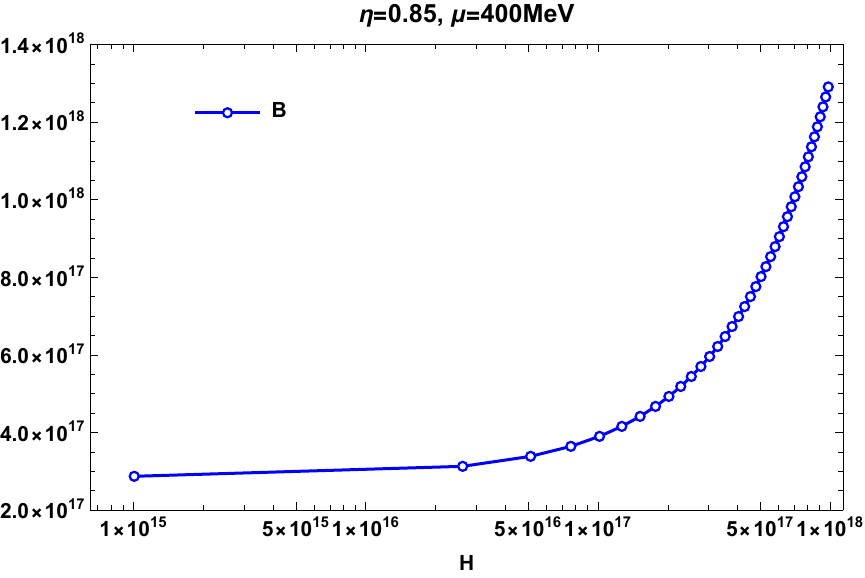} \caption{The induced field $\tilde{B}$ versus the applied field $\tilde{H}$ for a diquark coupling $\eta=0.85$ and baryonic chemical potential $\mu=400$ MeV. Both applied and induced fields are in Gauss. }
\end{figure}

\subsection{Gluon vortices and magnetic-field amplification}

In principle, we could solve the coupled equations (\ref{gluoncondensateequation}) and (\ref{inducedmagneticfield}) to obtain the inhomogeneous gluon condensate and the induced magnetic field in the medium. But those equations are involved nonlinear partial differential equations difficult to solve directly. Instead, we will use the constraint that the mass of the charged gluons must be less than twice of the color superconducting gap, since otherwise the gluon will decay into two quasiparticles. With this goal,
we first substitute  $2{\tilde e}|G|^2$ in (\ref{gluoncondensateequation}) by using  (\ref{inducedmagneticfield}) and transform the resulting equation to momentum space by using the following suitable transformation for charged scalar fields  \cite{Ritus-Trans}
\begin{equation}
G(p)=\int d^4x E_p(x)G(x)
\end{equation}
with
\begin{equation}
E_p(x)={\cal N}\exp (ip^0x^0+ip^2x^2)D_n(\rho)
\end{equation}
where $D_n(\rho)$ is the parabolic cylinder functions with argument $\rho=\sqrt{2|{\tilde e}{\tilde B}|}(x^1-p^2/{\tilde e}{\tilde B})$, ${\cal N}=(4\pi|{\tilde e}{\tilde B}|)^{1/4}/\sqrt{n!}$ is the normalization factor, and $n=0,1,2,\cdots$ denotes the Landau levels.  We, thus, obtain in momentum space
\begin{equation}
\left\{-\left[(p^0)^2-{\tilde e}{\tilde B}(2n+1) \right]-(2{\tilde e}{\tilde B}-m_M^2)-\frac{g^2({\tilde H}-{\tilde B})}{{\tilde e}}   \right\}G(p)=0
\end{equation}
We can then deduce the dispersion relation of the charged gluons
\begin{equation}
E_n^2={\tilde e}{\tilde B}(2n+1)-(2{\tilde e}{\tilde B}-m_M^2)-\frac{g^2({\tilde H}-{\tilde B})}{{\tilde e}} 
\end{equation}

The stability requires that the rest energy of the charged gluons must be smaller than twice the 
quasiparticle energy gap, i.e.,
\begin{equation}
E_{n=0}^2=m_M^2-{\tilde e}{\tilde B}-\frac{g^2({\tilde H}-{\tilde B})}{{\tilde e}}\leq(2\Delta)^2
\end{equation}
Therefore, the equality 
\begin{equation}
m_M^2-{\tilde e}{\tilde B}-\frac{g^2({\tilde H}-{\tilde B})}{{\tilde e}}=4\Delta^2\label{inducedfield}
\end{equation}
will give the induced field $\tilde B$ for each applied field $\tilde H$. From the solutions of the gap equation and the squared Meissner mass in Figs. \ref{Fig_1}-\ref{Cromo-M-Inst-2}, together with Eq.(\ref{inducedfield}), one can find the induced field $\tilde{B}$ as a function of the applied field $\tilde{H}$ as it is shown in Fig. 5.

From Fig. 5, we notice that the boosting effect is most noticeable at relatively low field ${\tilde H}$. On the other hand, decreasing the applied magnetic field below $10^{15}$ G the value of the induced field does not significantly change from the one obtained at $10^{15}$ G since for the baryonic chemical potential under consideration both applied field strengths can be considered as weak fields coinciding their results  in a good lead with the zero-field approximation. 

\section{Possible origin of the inner magnetic field of magnetars}

In this section we will discuss a possible mechanism that can serve to generate the inner magnetic field of magnetars. The subclass of neutron stars known as magnetars  \cite{Thompson}, where the AXPs and the SGR are included, exhibits magnetic fields that can reach values of the order of $10^{15}$ G on the star surface \cite{Kaspi}.
Even higher magnetic fields of the order of $10^{17}$ G are expected to be sustained in their interiors, as suggested by various recent observations (see, e.g., \cite{Rea} and references therein), as well as from theoretical estimations \cite{Theor-estimations} and according to some numerical simulations \cite{Uryu}.

One way to explain how the field amplification that occurs in magnetars is originated, the so-called dynamo mechanism \cite{Dynamo} is considered. This mechanism is responsible for the conversion of kinetic energy of an electrically conducting fluid into magnetic energy. Based on this idea, let us consider the energy equipartition between the star magnetic energy and the rotational kinetic energy which is given by
\begin{equation}\label{EPE}
\left (\frac{4}{3}\pi R^3\right ) \left(\frac{B^2}{8\pi}\right )=\frac{1}{2} I \omega^2=\frac{1}{5}MR^2\frac{4\pi^2}{P^2}
\end{equation}
where $M$ and $R$ are the  stellar mass and radius respectively, $I$ is the moment of inertia of a spinning sphere, $P$ is the stellar period of revolution and $B$ is the stellar magnetic field. 

\begin{figure}
	\includegraphics[height=8cm]{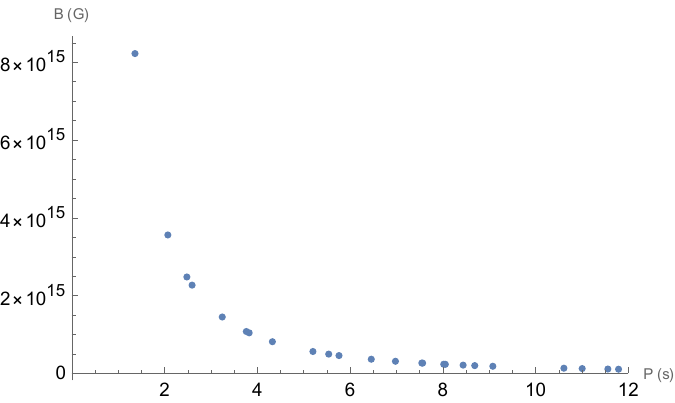}
		\caption{Induced magnetic field versus star spin period obtained by  considering the energy equipartition between the star magnetic energy and its rotational kinetic energy and taken  the spin periods from the  McGill Online Magnetar Catalog \cite{McGill}.}
	 \label{BvsP}
\end{figure}

If we write the energy equality equivalent to (\ref{EPE}) for the Sun and then divide (\ref{EPE}) by this second equation, we obtain 
\begin{equation}\label{EPE-Sun}
B=B_{\odot} \sqrt{\frac{MR_{\odot}}{M_{\odot} R}} \frac {P_{\odot}}{P}=\left ( \frac{0.8 \times 10^9} {P}  \right ) \textrm{G}
\end{equation}
with $M_{\odot}$, $R_{\odot}$ and $P_{\odot}$ being the solar mass, radius and rotational period respectively.
In (\ref{EPE-Sun}) we are taking the accepted values $M=1.4 M_{\odot}$ and $R=(0.14\times 10^{-4})R_\odot$. Regarding $P_{\odot}$, since the Sun is not a rigid body, it has different rotational periods depending on the altitude. We will consider here its minimum value of 25.67 days, which is the period at its equator, and that corresponds to  $P_{\odot}=2.2\times 10^6$ s.
Considering that $B_\odot=1$ G in Fig. \ref{BvsP} we plotted the magnetic field $B$ versus the period of rotation $P$ of magnetars obtained from (\ref{EPE-Sun}) with $P$ taken from the McGill Online Magnetar Catalog \cite{McGill}. 

The magnetic-field strengths found in Fig. \ref{BvsP} are of the order of those observed in  millisecond pulsars \cite {MillisecondPulsasr}, but they are far from the expected $10^{17}$ G inner field expected for magnetars.
Hence, we see that the order of magnitude of the magnetic field which corresponds to the macroscopic rotational kinetic energy of the star is not enough to induce the large inner field of the magnetars. This result suggests that the origin of this field could be associated with a microscopic mechanism that should be able to boost the inner field to larger values.

 We want to propose that precisely the mechanism we are describing in this paper can play this role. From Fig. 5 we see that if the core is formed by neutral 2SC color-superconducting quark matter in the chromomagnetic-unstable phase an applied magnetic field of  the order of $10^{15}$ G can be boosted to the expected value of $10^{17}$ G. Remarkably, this boosting mechanism works not only for moderately large applied magnetic field but also for fields with smaller magnitudes, say, of the order of $10^{8}$ G. 
 
  From (\ref{inducedfield}) 
\begin{equation}
{\tilde e}{\tilde B}\left(\frac{g^2}{{{\tilde e}}^2}-1\right)=\frac{g^2}{{{\tilde e}}^2}{\tilde e}{\tilde H}+4\Delta^2-m_M^2
\end{equation}

If the external field satisfies $\frac{g^2}{{{\tilde e}}}{\tilde H}<4\Delta^2-m_M^2$ with $-m_M^2>0$ , then 
\begin{equation}
{\tilde e}{\tilde B}\left(\frac{g^2}{{{\tilde e}}^2}-1\right)\simeq 4\Delta^2-m_M^2\label{approximateboostedfield}
\end{equation}
Thus, if the diquark coupling strength is not so large that makes the squared Meissner mass positive and comparable with $4\Delta^2$, the induced magnetic field  $\tilde B$ becomes of the order of the larger parameter between $\Delta$ and $m_M$.
Therefore the dependence of the boosted field on the external field exists only implicitly in the gap and Meissner mass, which, however, are almost constant for small and moderate external fields. Hence, the magnitude of an induced field $\tilde{B}$ for an applied field $\tilde{H}$ of the order of $10^{8}-10^9$ G will be essentially the same as that for $\tilde{H}$ of the order of $10^{15}$ G, from where we conclude that in the region with $\mu \in [320, 450]$ MeV and $1.15 > \eta > 0.75$ the induced field is of the order of $\tilde {B} \sim 10^{17}$ G as can be seen from Fig. 7 where Eq. (\ref{approximateboostedfield}) was used. There, we see that the expected value of $10^{17}$ G for the inner field of magnetars can be achieved by the proposed mechanism. We call attention that in Fig. 7 the results for $\eta = 1.15$ are not plotted. It is because at that coupling $m_M^2 > 0$ in the density region under consideration (see Figs. 3 and 4) and we need a strong-field approximation ($\tilde {e} \tilde{B} > m_M^2/2$) to produce the instability, which is a different situation than the one we are considering in the plots in Fig. 7.

\begin{figure}\label{Boosted_field_for_different_eta}
	\includegraphics[height=8cm]{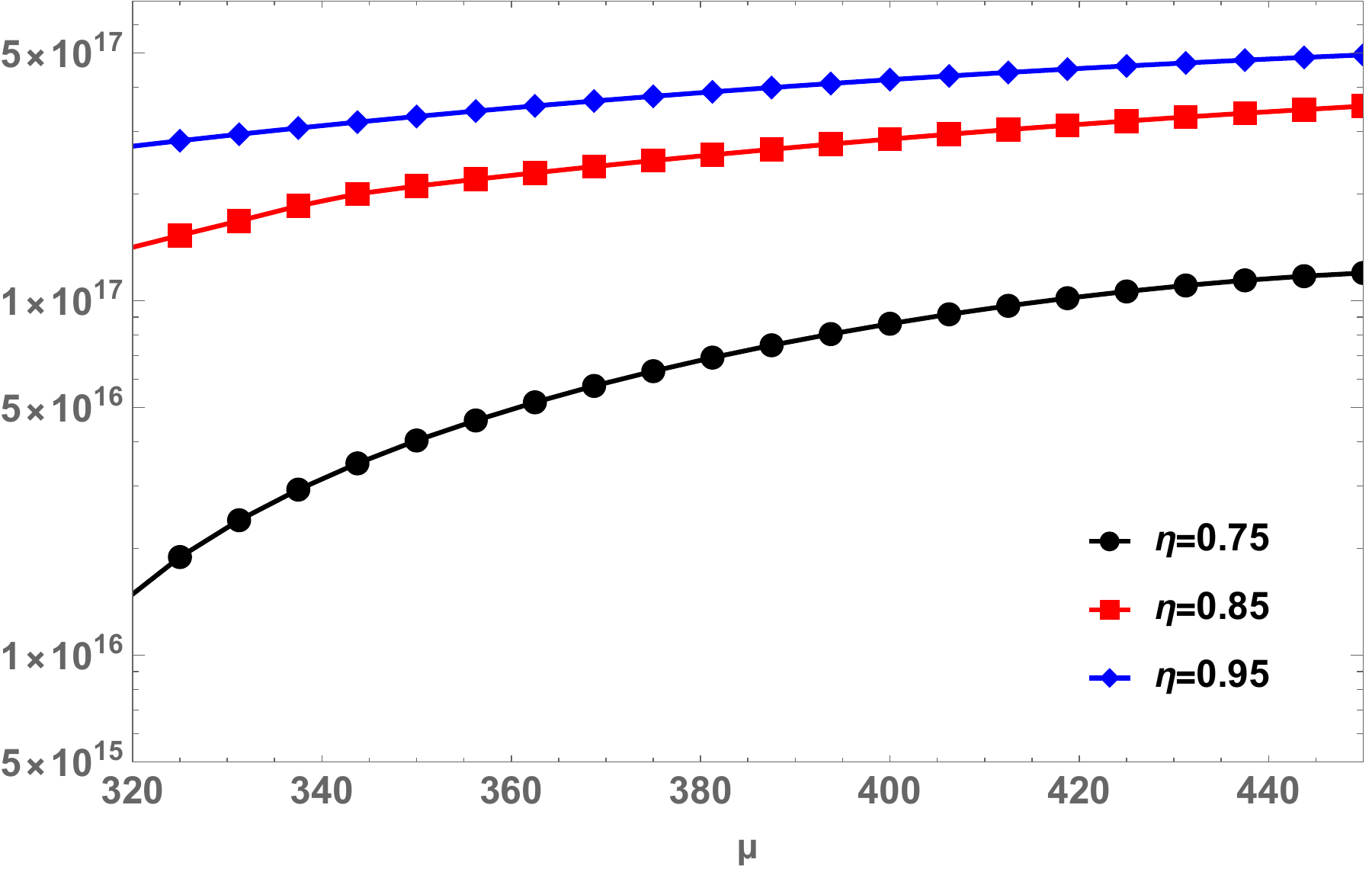} \caption{The induced field $\tilde{B}$ versus the chemical potential $\mu$ for different diquark couplings. The applied field is small enough so that we could use Eq.(\ref{approximateboostedfield}).}
\end{figure}

\section{Concluding remarks}

In this paper we attempt to find a mechanism that explains how the magnetar magnetic fields are generated.
Magnetars are compact astronomical objects endowed with very large magnetic fields. Using the observational data reported in the McGill Online Magnetar Catalog  \cite{McGill}, we calculated the magnitude of the generated stellar magnetic field by using the correlation between the star rotational energy and its magnetic energy. The found result was not sufficient to reproduce the expected field strength and suggests that the star field is not only produced by a dynamo effect of a charged rotational plasma, but for some qualitatively different inner mechanism.

Then, assuming that the star core is formed by neutral 2SC matter at intermediate baryonic densities we showed that at the densities of interest and for diquark coupling moderately large, the phase exhibits the so-called chromomagnetic instability with imaginary Meissner masses for the rotationally charged gluons. In this situation an applied magnetic field of any value will enforce the instability. This is the so-called "zero-mode problem" for spin-one charged fields in the presence of a magnetic field \cite{Ferrer-Vortices}-\cite{String Theory}. 

To remove the instability in this situation a restructuring of the ground state is proposed by the condensation of gluon vortices. This condensate, in addition to removing the instability, serves to boost the applied magnetic field. We showed that in the parameter region of interest a field of the order of $10^8$ G that could be generated by dynamo effect for the rotational frequencies reported for magnetars  \cite{McGill}, will be boosted to $10^{17}$ G by the mechanism here described. This value of $10^{17}$ G is the one indicated in several works as the most feasible one for the interior of magnetars \cite{Theor-estimations}. The lowest magnitude observed for the surface magnetic field  \cite{McGill} can be justified taking into account that in going from the core to the surface the magnetic field should decrease in order to satisfy the magnetic flux conservation in such a medium with a very high electric conductivity.

In this work, the findings about the  "zero-mode problem" for dense quark matter in the presence of a magnetic field  \cite{Ferrer-Vortices, MCFL-Vortices} have been applied to the neutral 2SC phase at ${ B}\ne 0$. Then, defining the diquark coupling and baryonic chemical potential in the expected parameter region for NS, we found the corresponding quantities as the gap, electric chemical potential and Meissner masses of charged gluons that will determine how much a weak field can be boosted by the found mechanism. Hence, we found numerically how much a weak magnetic field can be boosted under the given conditions of interest for magnetars.

\begin{acknowledgments}
The work of E. J. F. was supported in part by National Science Foundation Grant No. PHY-2013222 and Department of Energy Grant No. DE-SC0022023,  the work of A. P, was supported by DOE Grant No. DE-SC0022023 and the work of B. F. was supported by the National Natural Science Foundation of China (NSFC) under Grant No. 12075093.  E. J. F. is indebted to the Nuclear and Particle Astrophysics group at the Institute of Space Sciences (ICE-CSIC) and especially to Dr. Cristina Manuel, for warm hospitality. 
\end{acknowledgments}

\appendix

\newpage 

\end{document}